\documentclass[twocolumn]{aastex62}

\received{February 22, 2018}
\revised{August 30, 2018}
\accepted{September 4, 2018}
\submitjournal{ApJ}

\shorttitle{ICM properties of Einasto profile}
\shortauthors{M. S. Mirakhor}

\begin{document}

\title{PROPERTIES OF THE INTRACLUSTER MEDIUM ASSUMING AN EINASTO DARK
  MATTER PROFILE}

\email{mohammad.mirakhor@su.edu.krd}

\author{Mohammad S. Mirakhor}
\affil{Department of Physics, College of Science, Salahaddin
  University-Erbil\\
   Kirkuk Street, Erbil, 44001, Iraq}



\begin{abstract}

I investigate an analytical model of galaxy clusters based on the
assumptions that the intracluster medium plasma is polytropic and is in
hydrostatic equilibrium. The Einasto profile is adopted as a model for the
spatial-density distribution of dark matter halos. This model has
sufficient degrees of freedom to simultaneously fit X-ray surface
brightness and temperature profiles, with five parameters to describe the
global cluster properties and three additional parameters to describe
the cluster's cool-core feature. The model is tested with
{\it Chandra\/} X-ray data for seven galaxy clusters, including three
polytropic clusters and four cool-core clusters. It is found that the
model accurately reproduces the X-ray data over most of the radial
range. For all galaxy clusters, the data allows to show that the model
is essentially as good as that of \citeauthor{vikhlinin2006} and
\citeauthor{bulbul2010}, as inferred by the reduced $\chi^2$.

\end{abstract}



\section{Introduction} \label{sec:intro}

Galaxy clusters, the largest known objects with quasi-relaxed
structures in the Universe, are essential probes for tracing the growth
of cosmic structures and testing cosmological models
\citep{mantz2010,allen2011,hann2016}. Galaxies, ionized
intracluster medium (ICM), and dark matter are considered to be the main
components of galaxy clusters. The ICM, the diffuse plasma distributed
between galaxies in a cluster, accounts for most of the baryonic
constituent of galaxy clusters, with electron temperatures in the
approximate range of 1 to 10 keV. Studies of the ICM provide
insights into the formation and evolution of galaxy clusters
\citep[see e.g.][]{lin2012,barnes2016}. Furthermore, the ICM
yields valuable information on the structure and mass distribution of
galaxy clusters. For example, one can estimate the cluster's gas and total
masses from the ICM gas density and temperature profiles, under the
hypothesis of hydrostatic equilibrium, i.e. that the pressure exerted
by the gas is equally counterbalanced by the total gravitational
potential. From these mass measurements, it is possible to estimate
the gas mass fraction. This gas mass fraction can then be used, with a
constraint on the baryon matter density from big-bang nucleosynthesis
or Cosmic Microwave Background (CMB) measurements, to set a tight
constraint on the total matter density \citep[see
  e.g.][]{ettori2013,mantz2014,applegate2016}.

Throughout the past few decades, various models have been employed to
analyze the structure and morphology of the ICM plasma, most notably
the standard isothermal $\beta$-model \citep{cavaliere1976}. It
is found that the $\beta$-model provides a reasonable fit to the density
structure of the ICM plasma within relatively small radii
\citep[see e.g.][]{mohr1999,laroque2006}. However, this model
is limited in its ability to reproduce all observed features of the
density distribution of the ICM plasma. This is the case, for example,
in relaxed galaxy clusters with central-cuspy profiles
\citep[e.g.][]{pointecouteau2004}. Moreover, deep X-ray
observations with {\it Chandra\/} and {\it XMM-Newton\/} found that
the mean temperature profile declines on a large scale
\citep[see e.g.][]{vikhlinin2005,leccardi2008}.

Accordingly, other approaches have been attempted to circumvent these
issues. One such approach is to use the equation of state to relate the
pressure ($P$), density ($\rho$), and temperature ($T$) of the
ICM plasma in thermodynamic equilibrium, i.e. $f(P, \rho, T)=0$
\citep{horedt2004}. Under certain physical conditions, one finds that the
pressure and density are related by a polytropic equation of
state as $P \propto \rho ^{\gamma}$, where $\gamma$ is the polytropic index,
with values that are expected to range from 1 for an isothermal gas to
5/3 for an adiabatic gas. With the aid of the polytropic equation of state,
several studies introduced a revised version of the $\beta$-model to
describe the observable properties of galaxy clusters \citep[see
  e.g.][]{markevitch1999,ettori2000}. These studies found that the
mass measurements of typical galaxy clusters under the assumption of
the polytropic model can vary significantly within the virial regions,
when compared to those derived under the isothermal
assumption. However, \citet{degrandi2002} concluded that the
polytropic $\beta$-model does not provide a reasonable fit to the
temperature measurements over the full cluster-radial range. Instead,
the authors found that the temperature measurements can be better
modeled by a phenomenological broken power law.

To describe a broader range of the ICM-related features, different
modified versions of the $\beta$-model have been proposed to model the
distribution of the ICM plasma, most notably the \citet{vikhlinin2006} model.
These authors suggested a model with 17 free parameters to reconstruct the
observable properties of the ICM gas. According to this model, the
three-dimensional profile of the gas density is
\begin{eqnarray}
  n_e^2(r)=n_{e01}^2\frac{(r/r_{c1})^{-\alpha}}{(1+r^2/r_{c1}^2)^{3\beta_1-\alpha/2}}\frac{1}{(1+r^3/r_s^3)^{\epsilon/3}}\nonumber \\
  +\frac{n_{e02}^2}{(1+r^2/r_{c2}^2)^{3\beta_2}},
\label{Equ.: Density vikhlinin06}
\end{eqnarray}
while the gas temperature is given by
\begin{equation}
  T_e(r)=T_{e0}\frac{(r/r_t)^{-a}}{(1+r^b/r_t^b)^{c/b}}\tau_{\rm{cool}}(r),
  \label{Equ.: Temp vikhlinin06}
\end{equation}
where $\tau_{\rm{cool}}(r)$ is a phenomenological function that can be
expressed as \citep{allen2001}
\begin{equation}
  \tau_{\rm{cool}}(r)=\frac{\xi+(r/r_{\rm{cool}})^{a_{\rm{cool}}}}{1+(r/r_{\rm{cool}})^{a_{\rm{cool}}}},
  \label{Equ.: CC temperature}
\end{equation}
where $0 < \xi < 1$ is a free parameter, measuring the amount of
central cooling, $a_{\rm{cool}}$ is the shape parameter, and
$r_{\rm{cool}}$ is the cooling radius. Hence, 9 parameters ($n_{e01}$,
$r_{c1}$, $\beta_1$, $\alpha$, $r_s$, $\epsilon$, $n_{e02}$, $r_{c2}$,
and $\beta_2$) are used to describe the gas density, and 8 parameters
($T_{e0}$, $\xi$, $r_{cool}$, $a_{cool}$, $r_t$, $a$, $b$, and $c$)
for the gas temperature.

Although the \citet{vikhlinin2006} model provides a good fit to the
density and temperature profiles over the whole radial range, the
model is no longer considered physically motivated, but rather is
an ad-hoc model tailored to the observed data. Moreover, there are
many degeneracies between the model's best-fitting parameters,
implying less precise estimates for their values. This could cause a
major issue when one deals with relatively few data points or when one
attempts to extrapolate outside the range of observed data.

This has motivated more physically-grounded models with a limited
number of free parameters. \citet{ascasibar2008} presented a simple
analytical model based on the assumptions that the ICM plasma is
spherical symmetry and is in hydrostatic equilibrium. With only five
parameters, the authors found that the model can reconstruct the
gas density and temperature profiles of galaxy clusters yielded from
the best-fitting parameters of \citet{vikhlinin2006} with less than
20$\%$ discrepancy over most of the cluster-radial range.

More recently, \citet{bulbul2010} proposed an analytical model to
describe the observable properties of the diffuse ICM in galaxy
clusters based on the assumptions that the ICM plasma is polytropic
and is in hydrostatic equilibrium. In the polytropic state, the
electron number density, $n_e(r)$, and temperature, $T_e(r)$, of the
ICM gas are related using a simple power law \citep{ascasibar2003},
\begin{equation}
  \frac{n_{e}(r)}{n_{e0}}=\bigg[\frac{T_{e}(r)}{T_{e0}}\bigg]^n,
  \label{Equ.: Polytropic equation}
\end{equation}
where $n_{e0}$ and $T_{e0}$ are the central electron density
and temperature, respectively, and $n$ is the polytropic index. Under
the assumption of hydrostatic equilibrium, one can relate the
properties of the ICM gas to the total gravitational
  potential, $\phi(r)$,
\begin{equation}
  \frac{1}{\rho_e} \frac{dP_e(r)}{dr} = -\frac{d\phi(r)}{dr},
  \label{Equ.: HE}
\end{equation}
where $\rho_e\, (=\mu m_p n_e)$ is the electron
density, $\mu$ is the mean mass per particle in units of the
proton mass $m_p$, $P_e(r)\, (=n_e k_{\rm{B}} T_e)$ is the electron
pressure, and $k_{\rm{B}}$ is the Boltzmann constant.

From Equations (\ref{Equ.: Polytropic equation}) and (\ref{Equ.: HE}),
the polytropic temperature distribution can then be derived as a
function of the gravitational potential,
\begin{equation}
  T_{e}(r)=-\frac{1}{n+1}\frac{\mu m_{p}}{k_{\rm{B}}}\phi(r).
  \label{Equ.: T and GP}
\end{equation}
%
%
%

However, \citet{bulbul2010} dropped the polytropic assumption to account
for the gas cooling in the cluster center. Adopting a generalized form
of the \citet{navarro1996} profile (NFW), with density slope in
the outer regions controlled by a free parameter $\beta$, the
three-dimensional gas density and temperature profiles derived by
\citet{bulbul2010} are
\begin{equation}
  n_{e}(r)=n_{e0}\bigg(\frac{1}{(\beta-2)}\frac{(1+r/r_s)^{\beta-2}-1}{r/r_s(1+r/r_s)^{\beta-2}}\bigg)^n\tau_{\rm{cool}}^{-1}(r),
  \label{Equ.: Density bulbul}
\end{equation}
and
\begin{equation}
  T_{e}(r)=T_{e0}\bigg(\frac{1}{(\beta-2)}\frac{(1+r/r_s)^{\beta-2}-1}{r/r_s(1+r/r_s)^{\beta-2}}\bigg)\tau_{\rm{cool}}(r),
  \label{Equ.: Temp bulbul}
\end{equation}
respectively, with five free parameters ($n_{e0}$, $T_{e0}$, $r_s$,
$\beta$, and $n$) to describe global-cluster properties and three
additional parameters ($r_{\rm {cool}}$, $a_{\rm {cool}}$, and $\xi$)
to describe the cluster's cool-core feature.

In recent years, however, many observational studies
\citep[e.g.][]{silva2013,umetsu2014} and high-resolution {\it
  N\/}-body simulations
\cite[e.g.][]{hayashi2008,dhar2010,dutton2014,klypin2016}
have indicated that the Einasto profile \citep{navarro2004} provides a
better fit to the spatial-density distribution of dark matter halos
than does the NFW profile. In this paper, therefore, I adopt the
Einasto profile as a model for dark matter halos instead of the
generalized NFW model used by \citet{bulbul2010}. The model presented
in this work represents a slight variation in respect to the model
introduced by \citet{bulbul2010}. The Einasto profile, which has three
parameters, has a logarithmic slope that decreases inward more
gradually than the singular two-parameter profiles. Furthermore, this
three-parameter model allows the density profile to be tailored to
each individual halo, thereby yielding improved fits \citep[see
  e.g.][]{navarro2004}. Adopting this profile, I derive analytical
expressions for the thermodynamic properties of the ICM gas relevant
to X-ray observation. The model is tested with X-ray data of a sample
of seven galaxy clusters. All of the clusters have sufficient
signal-to-noise to enable accurate analysis for the radial profiles of
projected gas density and temperature. Moreover, the model is compared
with the \citet{vikhlinin2006} model (Equations (\ref{Equ.: Density
  vikhlinin06}) and (\ref{Equ.: Temp vikhlinin06})) and the
\citet{bulbul2010} model (Equations (\ref{Equ.: Density bulbul}) and
(\ref{Equ.: Temp bulbul})).

This paper is structured as follows: Section~\ref{sec:model} describes
the model; testing the model with X-ray {\it Chandra\/} data and
comparing it with previous analytical models are presented in
Section~\ref{sec:Test}. Finally, in
Section~\ref{sec:discussion}, I discuss and conclude the
results. Throughout this paper, I adopt a $\Lambda$ CDM cosmology with
$\Omega_{\Lambda}=0.7$, $\Omega_m=0.3$, and
$H_0=70 \,\rm{km}\,\rm{s}^{-1}\,\rm{Mpc}^{-1}$.

\section{MODELLING OF THE PROPERTIES OF THE INTRACLUSTER MEDIUM}
\label{sec:model}

\subsection{Einasto Profile}

The model developed in the current paper is essentially the same as
the one by \citet{bulbul2010} explained in Section~\ref{sec:intro},
but now the generalized NFW profile is replaced by the Einasto
profile. A generalized form of the Einasto profile for a spherical
density distribution at radius $r$ is
\begin{equation}
  \rho_{\rm{tot}}(r) = \rho_{-2}\exp\bigg\{-\frac{2}{\alpha}\bigg[\bigg(\frac{r}{r_{-2}}\bigg)^{\alpha}-1\bigg]\bigg\},
  \label{Equ.: The Einasto profile}
\end{equation}
where $r_{-2}$ and $\rho_{-2}$ are the radius and density at
which the logarithmic slope $d \ln \rho_{\rm{tot}}/d \ln r =  -2$,
and $\alpha$ is the shape parameter.

Since dark matter is the dominant component in galaxy clusters, the
Einasto profile is a good approximation for the total density
profile. This density profile is further combined with the polytropic
and hydrostatic equilibrium assumptions to derive the analytical
expressions for the total mass, the electron density, and the electron
temperature of the ICM gas.

\subsection{Total Mass and Potential Profiles}

The total mass of the galaxy cluster, which is mainly made
up of dark matter, can be determined by integrating the Einasto
profile \citep{retana-montenegro2012}
\begin{equation}
  M_{\rm{tot}}(r)=M_{0}\,\gamma\bigg(\frac{3}{\alpha},\frac{2 r^\alpha}{\alpha
    r_{-2}^\alpha}\bigg),
  \label{Equ.: Polytropic total mass}
\end{equation}
where $\gamma(a,x)$ is the lower incomplete gamma function,
and
\begin{equation}
  M_{0}=4\pi \rho_{-2} r_{-2}^3
  \frac{\alpha^{3/\alpha-1}}{2^{3/\alpha}}\exp\bigg(\frac{2}{\alpha}\bigg).
  \label{Equ.: Central total mass}
\end{equation}

\begin{figure}
\plotone{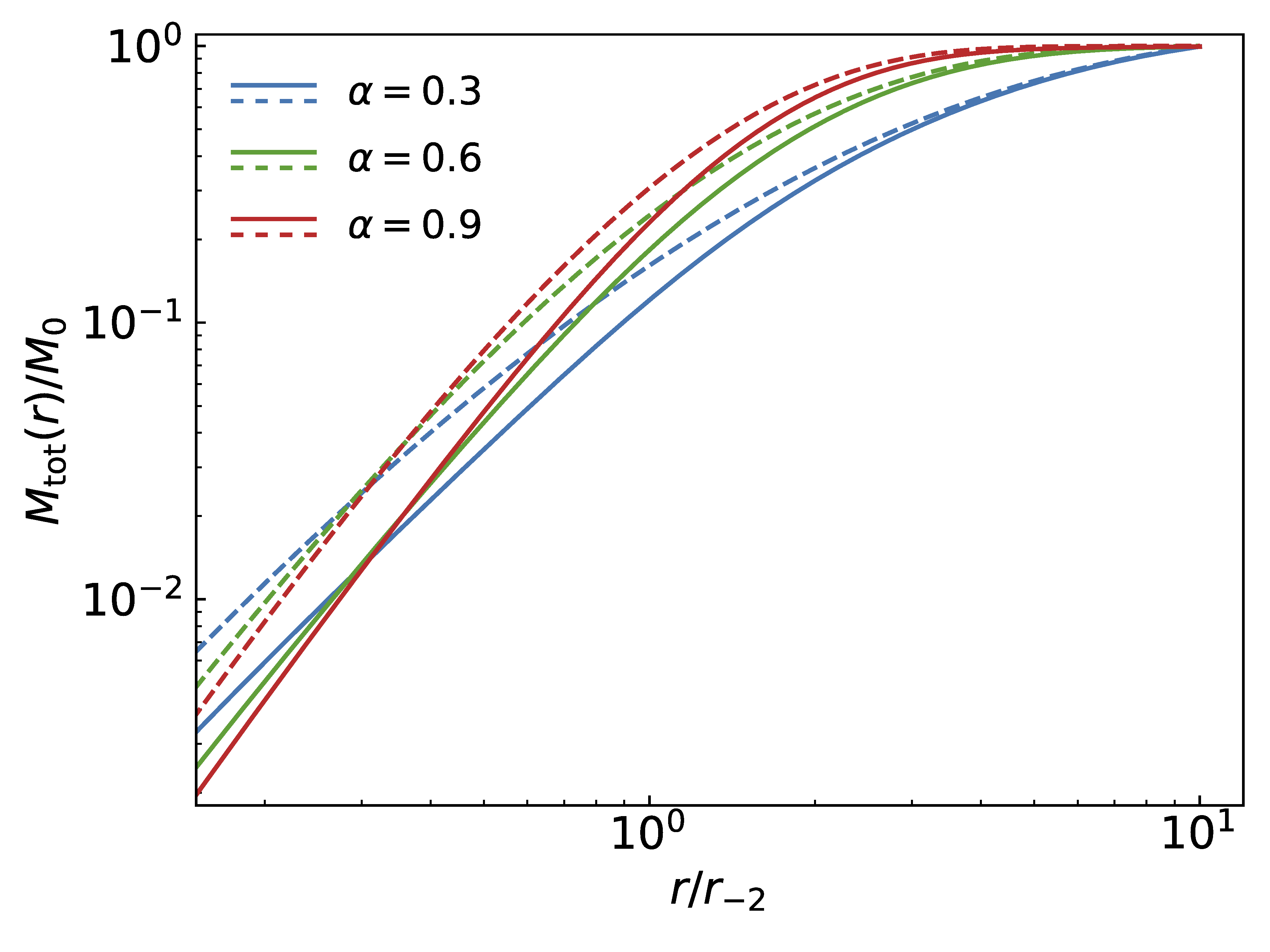}
\caption{Normalized total mass profiles with and
  without the phenomenological correction (Equation (\ref{Equ.: CC
    temperature})). Coloured dashed lines indicate the normalized total
  mass profiles for a phenomenological function with parameters of
  $\xi = 0.5$, $a_{\rm{cool}} = 2.0$, and $r_{\rm{cool}} = r_{-2}$;
  coloured solid lines indicate profiles without the phenomenological
  function. Total mass is finite and the cool-core correction is
  important only in the internal regions.}
\label{Fig.: M_tot}
\end{figure}

Figure \ref{Fig.: M_tot} shows the total mass profiles for various
values of the $\alpha$ parameter. The total mass profiles are finite since the
Einasto density profile cuts off exponentially at large radii.

For a spherically symmetric mass distribution, the cluster's gravitational
potential is related to the total gravitating mass by
\begin{equation}
  \phi(r) = \int\frac{G M_{\rm{tot}}(r)}{r^2} dr,
  \label{Equ.: GP}
\end{equation}
where $G$ is the Newtonian gravitational constant. Using Equation
(\ref {Equ.: Polytropic total mass}), the gravitational potential can
be found as \citep{retana-montenegro2012}
\begin{eqnarray}
 \phi(r)=\phi_0 \bigg[\bigg(\frac{2 r^\alpha}{\alpha
     r_{-2}^\alpha}\bigg)^{-1/\alpha} \gamma\bigg(\frac{3}{\alpha},
   \frac{2 r^\alpha}{\alpha r_{-2}^\alpha}\bigg) \nonumber \\
   +  \Gamma\bigg(\frac{2}{\alpha},\frac{2 r^\alpha}{\alpha r_{-2}^\alpha}\bigg) \bigg],
 \label{Equ.: GP solution}
\end{eqnarray}
where
\begin{equation}
 \phi_0=-4\pi G \rho_{-2} r_{-2}^2
 \frac{\alpha^{2/\alpha-1}}{2^{2/\alpha}}\exp\bigg(\frac{2}{\alpha}\bigg),
 \label{Equ.: Central GP}
\end{equation}
and $\Gamma(b,x)$ is the upper incomplete gamma function, which is
related to the lower incomplete gamma function, $\gamma(b,x)$,
through the complete gamma function,
\begin{equation}
  \Gamma(b)=\Gamma(b,x)+\gamma(b,x).
  \label{Equ.: Gamma relation}
\end{equation}

\subsection{Temperature and Density Profiles}

Using Equations (\ref{Equ.: T and GP}) and (\ref{Equ.: GP solution}),
the polytropic temperature profile becomes
\begin{eqnarray}
  T_{e,\rm{poly}}(r)=T_{e0} \bigg[\bigg(\frac{2 r^\alpha}{\alpha
     r_{-2}^\alpha}\bigg)^{-1/\alpha}
  \gamma\bigg(\frac{3}{\alpha},
  \frac{2 r^\alpha}{\alpha r_{-2}^\alpha}\bigg) \nonumber \\
  + \Gamma\bigg(\frac{2}{\alpha},\frac{2 r^\alpha}{\alpha
       r_{-2}^\alpha}\bigg) \bigg],
  \label{Equ.: Temperature profile}
\end{eqnarray}
where
\begin{equation}
  T_{e0}=\frac{4\pi G \rho_{-2} r_{-2}^2}{n+1}\frac{\mu m_{p}}{k_{\rm{B}}}
  \frac{\alpha^{2/\alpha-1}}{2^{2/\alpha}}\exp\bigg(\frac{2}{\alpha}\bigg).
  \label{Equ.: Central T}
\end{equation}

Taking advantage of the power-law relation between the density and
temperature of the ICM plasma (Equation (\ref{Equ.: Polytropic
  equation})), the polytropic electron density is
\begin{eqnarray}
  n_{e,\rm{poly}}(r)=n_{e0}\bigg[\bigg(\frac{2 r^\alpha}{\alpha
     r_{-2}^\alpha}\bigg)^{-1/\alpha}
  \gamma\bigg(\frac{3}{\alpha},
  \frac{2 r^\alpha}{\alpha r_{-2}^\alpha}\bigg) \nonumber \\
  + \Gamma\bigg(\frac{2}{\alpha},\frac{2 r^\alpha}{\alpha
       r_{-2}^\alpha}\bigg) \bigg]^n.
  \label{Equ.: Electron density}
\end{eqnarray}

\subsection{Pressure profile}

Using the ideal gas law expression, $P_e(r)=n_e k_{\rm{B}}T_e$,
and with the help of Equations (\ref{Equ.: Temperature profile}) and
(\ref{Equ.: Electron density}), the pressure profile for polytropic
clusters is given by
\begin{eqnarray}
  P_{e}(r)=P_{e0} \bigg[\bigg(\frac{2 r^\alpha}{\alpha
     r_{-2}^\alpha}\bigg)^{-1/\alpha}
  \gamma\bigg(\frac{3}{\alpha},
  \frac{2 r^\alpha}{\alpha r_{-2}^\alpha}\bigg) \nonumber \\
  +  \Gamma\bigg(\frac{2}{\alpha},\frac{2 r^\alpha}{\alpha
       r_{-2}^\alpha}\bigg) \bigg]^{n+1},
  \label{Equ.: Pressure}
\end{eqnarray}
where $P_{e0}\,(=n_{e0}k_{\rm{B}}T_{e0})$ is the central pressure.

\subsection{Cool-Core Component}

Several studies have shown that the polytropic model can also be applied to
cool-core clusters, after adopting a cool-core-corrected temperature
profile \citep[see e.g.][]{landry2013}. The temperature profile
of such clusters that feature a decline in temperature in the central
region can be parameterized by
\begin{eqnarray}
  T_{e,
    \rm{cool}}(r)=T_{e0} \bigg[\bigg(\frac{2 r^\alpha}{\alpha
     r_{-2}^\alpha}\bigg)^{-1/\alpha}
  \gamma\bigg(\frac{3}{\alpha},
  \frac{2 r^\alpha}{\alpha r_{-2}^\alpha}\bigg) \nonumber \\
  +  \Gamma\bigg(\frac{2}{\alpha},\frac{2 r^\alpha}{\alpha
       r_{-2}^\alpha}\bigg) \bigg] \tau_{\rm{cool}}(r).
  \label{Equ.: Modified temperature profile}
\end{eqnarray}

Recent X-ray and SZ effect observations \citep[see
  e.g.][]{arnaud2010,sayers2013} have found that the ICM pressure
profile, when scaled appropriately, follows a nearly universal shape,
suggesting that it is relatively independent of morphology and
dynamical state of the ICM gas. Accordingly, following
\citet{bulbul2010}, in this work it is assumed that the pressure
distribution is the same for polytropic and cool-core clusters
(Equation (\ref{Equ.: Pressure})). The density profile for cool-core
  clusters, therefore, can be obtained using
  $n_{e,\rm{cool}}(r)=P_e(r)/k_{\rm{B}}T_{e,\rm{cool}}(r)$,
\begin{eqnarray}
  n_{e,\rm{cool}}(r)
             =n_{e0}\bigg[\bigg(\frac{2 r^\alpha}{\alpha
     r_{-2}^\alpha}\bigg)^{-1/\alpha}
  \gamma\bigg(\frac{3}{\alpha},
  \frac{2 r^\alpha}{\alpha r_{-2}^\alpha}\bigg) \nonumber \\
  +  \Gamma\bigg(\frac{2}{\alpha},\frac{2 r^\alpha}{\alpha
       r_{-2}^\alpha}\bigg) \bigg]^n\tau_{\rm{cool}}^{-1}(r).
  \label{Equ.: Modified electron density}
\end{eqnarray}

To keep the hydrostatic equilibrium assumption (Equation (\ref{Equ.:
  HE})), these modified temperature (Equation (\ref{Equ.: Modified
  temperature profile})) and density (Equation (\ref{Equ.: Modified
  electron density})) profiles require to introduce a modified version
for the total mass enclosed in radius $r$,
\begin{equation}
  M_{\rm{tot}}(r)=M_{0}\,\gamma\bigg(\frac{3}{\alpha},\frac{2 r^\alpha}{\alpha
    r_{-2}^\alpha}\bigg) \tau_{\rm{cool}}(r),
  \label{Equ.: CC total mass}
\end{equation}
where the $\tau_{\rm{cool}}(r)$ term is significant only in the central
region (see Figure \ref{Fig.: M_tot}).

\subsection{Surface Brightness Profile}

The X-ray surface brightness, $S_{\rm{X}}(R)$ of a galaxy cluster
along the line of sight, $dl$, is related to the electron density and
temperature distributions of the ICM gas by \citep{suto1998}
\begin{equation}
  S_{\rm{X}}(R)=\frac{1}{4\pi(1+z)^4}\int n_{e}^2(r)\Lambda_{ee}(T_{e}) dl,
  \label{Equ.: The surface brightness}
\end{equation}
where $z$ is the cluster redshift, and $\Lambda_{ee}(T_{e})$ is the
X-ray spectral emissivity, which depends on temperature and
metallicity, and is calculated in units of counts cm$^3$ s$^{-1}$
using the average cluster temperature. Substituting Equations
(\ref{Equ.: Modified temperature profile}) and (\ref{Equ.: Modified
  electron density}) into Equation (\ref{Equ.: The surface
  brightness}), the radial profiles of the X-ray surface brightness
profile can then be obtained by numerical integration.

\begin{figure}
\plotone{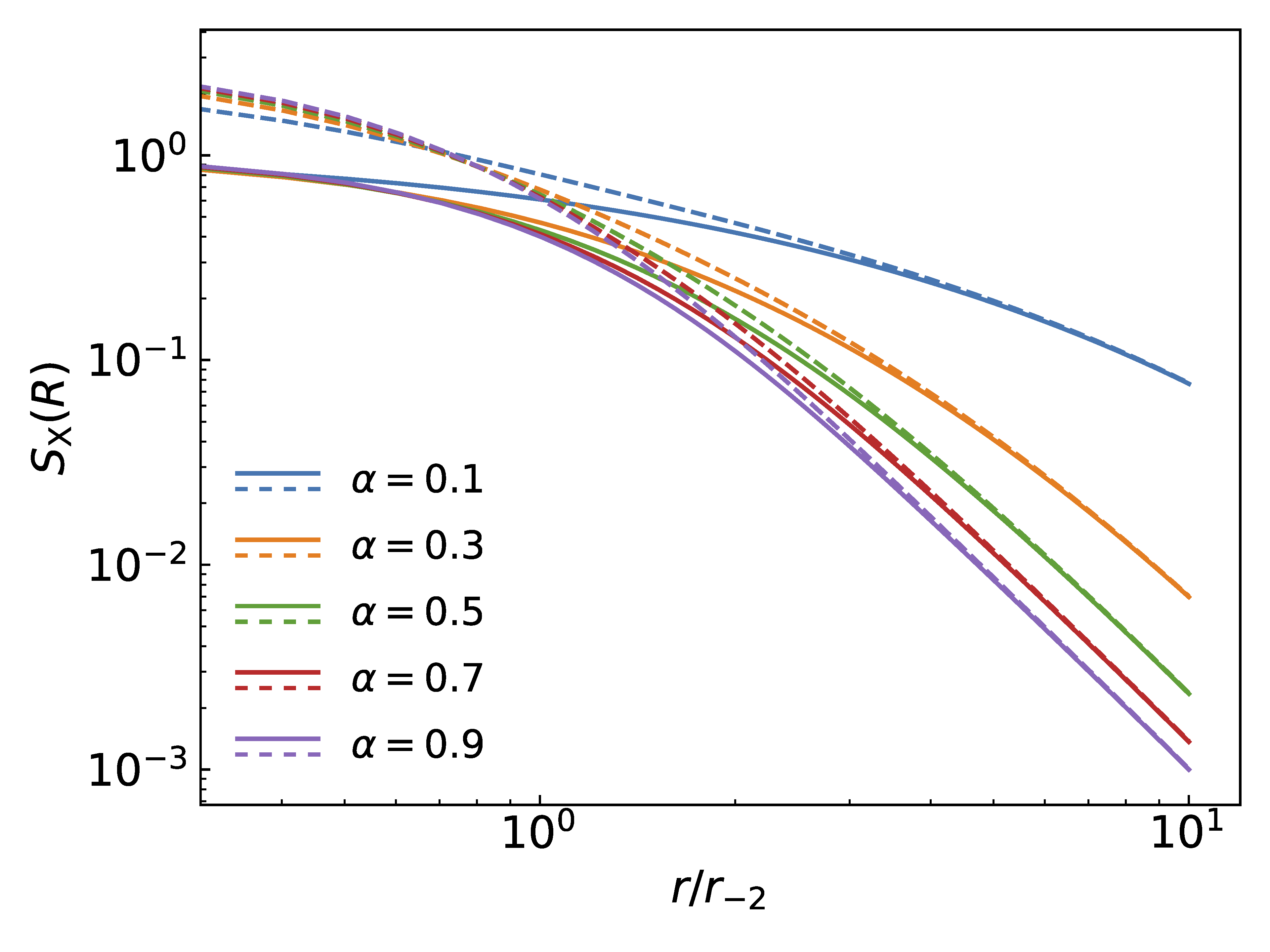}
\caption{Normalized X-ray surface brightness profiles with and
  without the phenomenological correction (Equation (\ref{Equ.: CC
    temperature})) for different values of the $\alpha$ parameter. The
  quantity $S_{X}(R)$ is shown in arbitrary units. Coloured dashed
  lines indicate the normalized X-ray surface brightness profiles for
  a phenomenological function with parameters of $\xi = 0.5$,
  $a_{\rm{cool}} = 2.0$, $r_{\rm{cool}} = r_{-2}$, and $n =
  2.0$; coloured solid lines indicate profiles without the
  phenomenological function. The effect of the phenomenological
  function is significant only in the central region.}
\label{Fig.: Sx}
\end{figure}

Figure \ref{Fig.: Sx} depicts the radial profiles of the X-ray surface
brightness with and without the phenomenological function for
different values of the $\alpha$ parameter. The effect of the
phenomenological function is observable only in the cluster's central
region.

\section{TESTING MODEL WITH X-RAY OBSERVATIONS}
\label{sec:Test}
\subsection{Data Sample}
\label{sec:data sample}

To test the model, archival {\it Chandra\/} X-ray data of seven galaxy
clusters were analyzed.  The source name, the {\it Chandra\/}
observation identification number, the exposure time, the redshift,
the Galactic absorption, and the redshift reference of the cluster
sample are listed in Table \ref{Table: Cluster data}. The selected
clusters have X-ray luminosities in the 0.1$-$2.4 keV band of
$L_{\rm{X,keV}} > 4 \times 10^{44}$ erg s$^{-1}$, with redshifts $0.10
< z < 0.60$. These galaxy clusters are selected since they have a
regular X-ray morphology, and show no or only weak signs of dynamical
activity, and the images have sufficient signal-to-noise to enable
accurate analysis for the radial profiles of the projected temperature
and X-ray surface brightness. Many of these clusters have been studied
in literature. In Section \ref{sec: mass comparison}, I compare the
mass measurements for Abell 1835, estimated from the current work,
with those reported in \citet{landry2013}.

\label{sec:observation}
\subsection{Data Reduction}

Data reduction was performed using the Chandra Interactive Analysis of
Observations (\textsc{ciao}) version 4.9, with the latest calibration
database (\textsc{caldb}) version 4.7.3. I reprocessed the {\it
  Chandra\/} data using the \texttt{\detokenize{chandra_repro}}
routine to perform the recommended data preparation, such as checking
the source coordinate, filtering the event file to good time
intervals, removing streak events, and identifying the bad
pixels. This script generates an event file and a bad-pixel
file. Since all observations were taken in VFAINT mode, events with
significant positive pixels at the border of the event island were
excluded by further filtering.

\begin{deluxetable*}{lCCCCC}
\tablenum{1}
\tablecaption{Cluster data.  From left to
right, columns give the source name, the {\it Chandra\/}
observation identification number, the exposure time, the redshift,
the Galactic absorption (from \citet{dickey1990}), and the redshift
reference.\label{Table: Cluster data}}
\tablewidth{0pt}
\tablehead{
\colhead{Cluster} &
\colhead{ObsID} &
\colhead{Exposure} &
\colhead{$z$} &
\colhead{$N_{\textrm{H}}$} &
\colhead{$z$ Reference}\\
\colhead{} &
\colhead{} &
\colhead{(ks)} &
\colhead{} &
\colhead{($10^{20}$ cm$^{-2}$)} &
\colhead{}
}
\startdata
Abell 2218        & 1666  & 40.7  & 0.176 & 2.57 & \text{\citet{struble1999}} \\
Abell 1835        & 6880  & 117.1 & 0.253 & 2.04 & \text{\citet{struble1999}} \\
Abell 2050        & 18251 & 14.9  & 0.120 & 4.66 & \text{\citet{ebeling1996}} \\
Abell 1689        & 7289  & 74.9  & 0.183 & 1.81 & \text{\citet{struble1999}} \\
MACS J0647.7+7015 & 3196  & 19.1  & 0.584 & 5.24 & \text{\citet{laroque2003}} \\
MACS J1423.8+2404 & 4195  & 114.2 & 0.545 & 2.45 & \text{\citet{laroque2003}} \\
RXC J2014.8-2430  & 11757 & 19.7  & 0.161 & 7.70 & \text{\citet{bohringer2004}} \\
\enddata
\end{deluxetable*}

As part of the data reduction, background light curves were examined
to detect and remove the flaring periods. For the background dataset,
the light curve is generated in the 0.3$-$12.0 keV energy band,
following the recommendations given by \citet{markevitch2003}. The
light curve is analysed using the \texttt{\detokenize{lc_sigma_clip}}
routine provided by the \textsc{python} script
\texttt{\detokenize{lightcurves.py}}. This routine removes data points
that lie outside a certain sigma value from the mean count rate (a
$3\sigma$ clip was used in this work).

\subsection{Background Subtraction}

An important aspect of analysis of extended objects is the background
subtraction. For this purpose, blank-sky backgrounds were extracted for all
observations, processed and reprojected onto the sky to match the
cluster observation. Although the background spectrum is remarkably
stable, there are short-term and secular changes of the background
intensity of as much as $30\%$ because of charged-particle
events. Following a method similar to that described in
\citet{vikhlinin2005}, small adjustments to the background
normalisation were applied to increase the accuracy of the
background, based on data in the energy range 9.5$-$12.0 keV, where
the effective area of the ACIS detector and the source
emission are almost zero.

Besides charged-particle events, soft X-ray emission contributes to
the blank-sky-background data. This soft background, which is likely
to arise from differences in the extragalactic and Galactic foreground
emissions between the source and blank-sky observations, was fitted
to a thermal-plasma model. The soft background model was then scaled
to the cluster-sky area, and was included as a fixed background
component in the cluster's spectral analysis.

\subsection{X-Ray Images and Spectra}

The X-ray images of the selected clusters were created in the energy
range of 0.5$-$7.0 keV in order to determine the radial surface brightness
profiles. This energy range was selected to minimise the
high-energy-particle background, which rises significantly at low and
high energies. An exposure-corrected image of the cluster's selected region
was created, and passed to the source-detection tool to detect and
remove point sources and extended substructures. Count values in the
point-source regions were replaced with those interpolated from their
background. The surface brightness profiles were then extracted in
concentric annuli centred at the {\it Chandra\/} selected
centre.

Like the surface brightness profiles, X-ray spectra were also
extracted in the energy range of 0.5$-$7.0 keV from concentric annuli
centred at the {\it Chandra\/} selected centre, after excluding point
sources and extended substructures. The X-ray spectrum of
every galaxy cluster was fitted to a MEKAL model \citep{mewe1985},
modified by local Galactic absorption. The temperature, abundance,
and normalisation were left free in all annuli, whereas the
cluster redshift, $z$, and the Galactic absorption, $N_{\rm{H}}$, were
fixed (see Table \ref{Table: Cluster data}).

\subsection{Sources of Systematic Uncertainty}
\label{subsec: systematic}

The radial surface brightness and temperature profiles are
subject to various sources of systematic error. The choice
of the local background region is the major source of error
in the calibration of the {\it Chandra\/} observations. I follow
\citet{bulbul2010} and consider a $\pm 5\%$ uncertainty in the
imaging- and spectral-data analysis due to the variation of the
count rates of different background regions. Contamination on the
optical filter is another major source of uncertainty. This affects
the spectral data by $5 \%$ \citep{bulbul2010}, and is added in
quadrature to each energy bin in the temperature data. Another
possible source of uncertainty is the spatially dependent
non-uniformity in the effective area of the ACIS detector. The spatial
dependence of the detector efficiency scatters at a level of $\pm 1\%$
\citep{bulbul2010}. This uncertainty, also, is added in quadrature to
each annulus in the count rates.

\subsection{Model Fitting}

A Monte Carlo Markov Chain (MCMC) approach is adopted, as illustrated
in \citet{bonamente2004}, for the fitting process. The parameter space in
this approach is explored by moving randomly from a set of parameters
to another using the Metropolis-Hastings algorithm. This algorithm
typically accepts a move to the new point with the likelihood higher
than the old one. Hence, the algorithm gradually moves towards the
highest likelihood regions, where the parameter values yield the best
fit to the data.

The MCMC method is used to independently calculate the likelihood of
the spatial and spectral data with the model. After binning the X-ray
data, the log likelihood for the spatial data is given by
\begin{equation}
  \ln(\mathcal{L}_{\rm{spatial}})=-\frac{1}{2}\chi^2-\frac{1}{2}\sum_i \ln(2\pi\sigma_i^2),
  \label{Equ.: The spatial likelihood}
\end{equation}
where $\chi^2=\sum_i[(D_i-M_i)/\sigma_i]^2$, $D_i$ is the number of
counts detected in bin $i$, $M_i$ is the number of counts predicted by
the model in bin $i$, and $\sigma_i$ is the measured uncertainty on
$D_i$.

For the spectral data, the log likelihood,
$\ln(\mathcal{L}_{\rm{spectral}})$, is the same as in the spatial case
(Equation (\ref{Equ.: The spatial likelihood})), except here
$\chi^2=\sum_i[(T_i-M_i)/\sigma_i]^2$, where $T_i$ and $M_i$ are the
measured and predicted temperatures, respectively, and $\sigma_i$ is
the measured uncertainty on $T_i$. Then, the joint likelihood of the spatial
and spectral models
($\mathcal{L}=\mathcal{L}_{\rm{spatial}}\mathcal{L}_{\rm{spectral}}$)
is calculated, and the goodness of fit is tested using the $\chi^2$
statistic.

Adopting the MCMC approach, radial profiles of temperature and
background-subtracted surface brightness are fitted to the model
(Equations (\ref{Equ.: Modified temperature profile}) and (\ref{Equ.:
  The surface brightness})). Five parameters are used to model the
global-cluster properties, whereas three describe the cluster's
central region. At the beginning, all clusters were fit to the model
letting all parameters, including the cool-core parameters, free to
vary. For three clusters (Abell 2218, Abell 2050, and MACS
J0647.7+7015), however, it is found that the shape parameter $a_{cool}
\approx 1$, suggesting that there is no need for the cool-core
parameters for these clusters. For these three clusters, therefore,
the $a_{cool}$ parameter is set to 1, and then the X-ray data fitted to
the model, i.e. using only the global parameters. I classified such
clusters as polytopic clusters, i.e. do not possess a cool-core
component, whereas the remaining clusters are classified as cool-core
clusters.

\begin{figure*}
\plotone{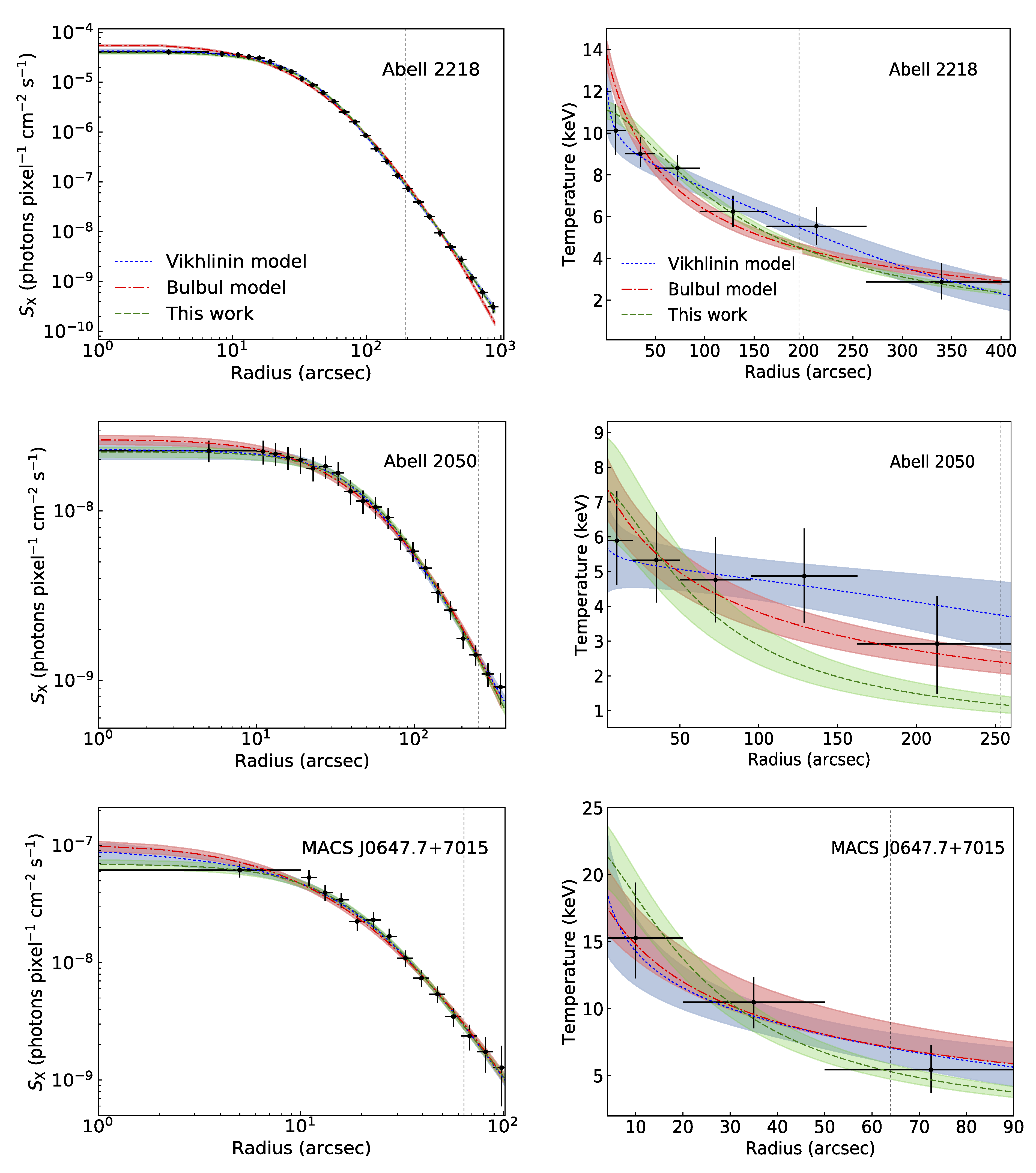}
\caption{Background-subtracted X-ray surface brightness (left panel) and
  temperature (right panel) distributions for polytropic clusters. In all profiles,
  the green dashed line indicates the best-fitting model proposed by
  this work. For comparison, the red dot-dash line indicates the
  best-fitting \citet{bulbul2010} model, and the blue dotted line
  indicates the best-fitting \citet{vikhlinin2006} model. The shadow
  regions indicate the 68.3$\%$ confidence intervals obtained by MCMC
  simulations, and the vertical-dashed line indicates the $r_{2500}$ radius.}
\label{Fig.: Polytropic clusters}
\end{figure*}

\begin{figure*}
\plotone{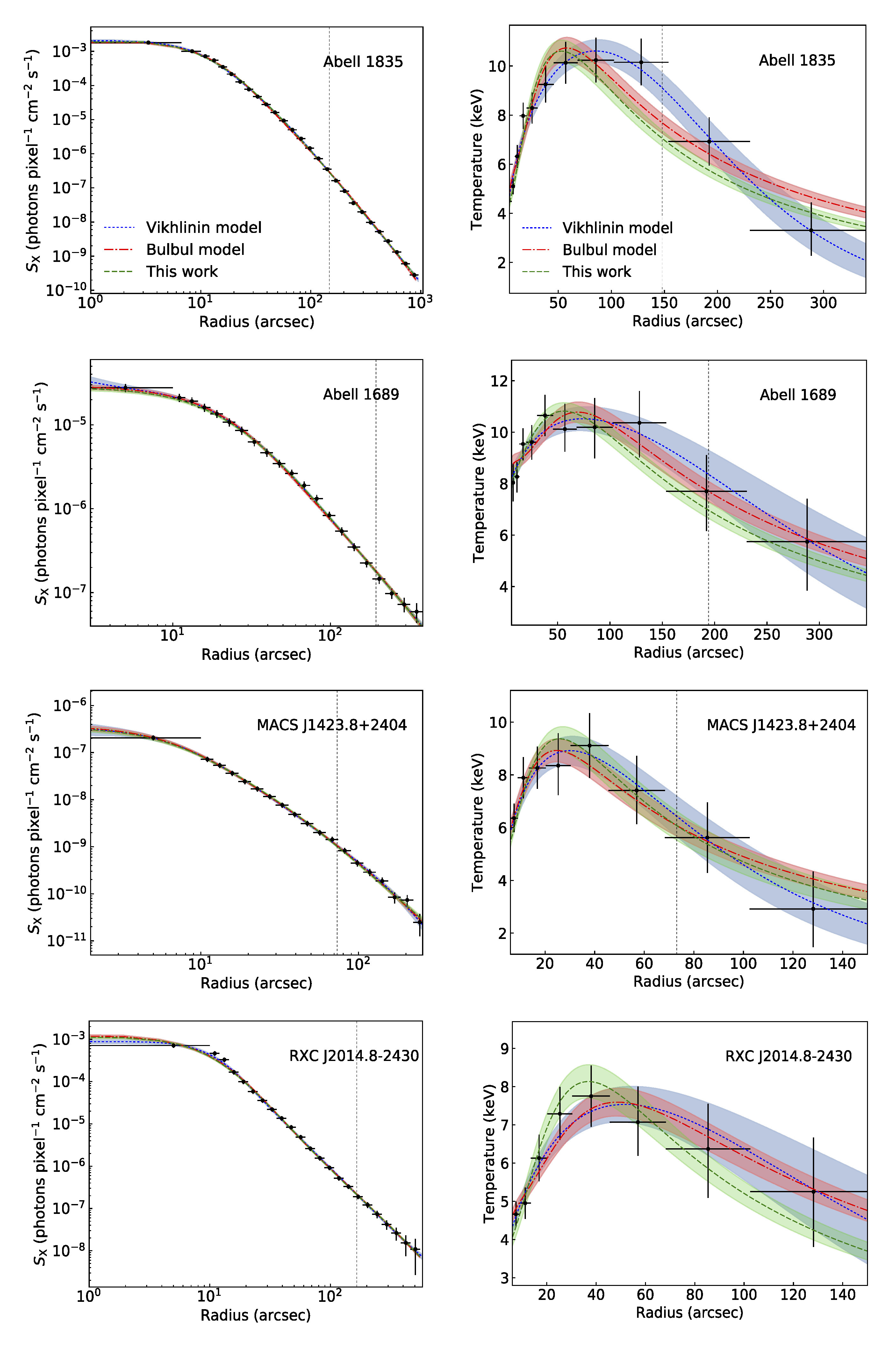}
\caption{Same as Figure~\ref{Fig.: Polytropic clusters}, except for
  cool-core clusters.}
\label{Fig.: cc clusters}
\end{figure*}

The values of the best-fitting parameters for the gas temperature
(Equation (\ref{Equ.: Modified temperature profile})) and density
(Equation (\ref{Equ.: Modified electron density})) profiles are listed
in Table \ref{Table: Best-fitting parameters} (see Table \ref{Table:
  Comparison} for their associated reduced $\chi_{\rm{tot}}^2$). The
best-fitting surface brightness and temperature profiles of all
clusters are presented in Figures \ref{Fig.: Polytropic clusters} and
\ref{Fig.: cc clusters}. For most polytropic and cool-core clusters,
the model accurately reproduces the X-ray surface brightness and
temperature profiles over the most radial range. For the Abell 1835
and Abell 2050 clusters, however, the model does not fit well the
temperature profile, particularly at intermediate to large radii.

\begin{deluxetable*}{lCCCCCCCC}
\tablenum{2}
\tablecaption{The model best-fitting parameters for the gas
  temperature (Equation (\ref{Equ.: Modified temperature profile}))
  and density (Equation (\ref{Equ.: Modified electron density}))
  profiles. Columns: (1) Cluster name; (2-6) Global parameters;
  (7-9) Cool-core parameters.
\label{Table: Best-fitting parameters}}
\tablewidth{0pt}
\tablehead{
\colhead{Cluster} &
\colhead{$n_{e0}$} &
\colhead{$r_{-2}$} &
\colhead{$\alpha$} &
\colhead{$n$} &
\colhead{$T_{e0}$} &
\colhead{$\xi$} &
\colhead{$r_{\rm cool}$} &
\colhead{$a_{\rm cool}$} \\
\colhead{} &
\colhead{($10^{-4}\,\rm cm^{-3}$)} &
\colhead{(arcsec)} &
\colhead{} &
\colhead{} &
\colhead{(keV)} &
\colhead{} &
\colhead{(arcsec)} &
\colhead{}
}
\startdata
Abell 2218       & $1.11_{-0.03}^{+0.02}$ & $78.41_{-1.70}^{+1.76}$ &
    $0.71_{-0.02}^{+0.02}$ & $2.44_{-0.02}^{+0.02}$ &
    $6.47_{-0.40}^{+0.42}$ & $-$ & $-$ & $-$  \\
Abell 1835       & $0.79_{-0.01}^{+0.01}$ & $24.52_{-1.07}^{+1.08}$ &
    $0.44_{-0.01}^{+0.02}$ & $2.39_{-0.04}^{+0.04}$ &
    $3.23_{-0.74}^{+0.82}$ & $0.11_{-0.01}^{+0.01}$ &
$47.61_{-5.18}^{+4.89}$ & $1.69_{-0.09}^{+0.12}$   \\
Abell 2050      & $5.12_{-0.34}^{+0.36}$ & $38.48_{-5.46}^{+6.84}$ &
$0.87_{-0.12}^{+0.09}$ & $1.20_{-0.06}^{+0.07}$ &
$6.19_{-1.44}^{+1.29}$ & $-$ & $-$ & $-$ \\
Abell 1689      & $4.35_{-0.18}^{+0.19}$ & $31.34_{-3.17}^{+3.30}$ &
$0.52_{-0.03}^{+0.03}$ & $1.36_{-0.08}^{+0.10}$ &
$9.45_{-2.39}^{+1.71}$ & $0.17_{-0.03}^{+0.04}$ &
$85.89_{-15.90}^{+19.50}$ & $1.49_{-0.15}^{+0.20}$  \\
MACS J0647.7+7015 & $54.21_{-2.33}^{+2.48}$ & $14.69_{-2.12}^{+2.45}$ &
$0.86_{-0.13}^{+0.08}$ & $1.60_{-0.09}^{+0.09}$ &
$17.91_{-4.39}^{+3.48}$ & $-$ & $-$ & $-$  \\
MACS J1423.8+2404 & $8.31_{-0.32}^{+0.34}$ & $17.78_{-1.87}^{+1.88}$ &
$0.56_{-0.03}^{+0.04}$ & $1.91_{-0.10}^{+0.12}$ &
$6.30_{-1.78}^{+2.14}$ & $0.17_{-0.05}^{+0.05}$ &
$20.10_{-3.27}^{+5.76}$ & $1.71_{-0.33}^{+0.38}$  \\
RXC J2014.8-2430 & $1.55_{-0.11}^{+0.12}$ & $18.18_{-1.03}^{+1.30}$ &
$0.66_{-0.05}^{+0.03}$ & $1.76_{-0.05}^{+0.05}$ &
$14.22_{-2.75}^{+2.21}$ & $0.12_{-0.02}^{+0.02}$ &
$38.71_{-4.68}^{+5.22}$ & $1.86_{-0.18}^{+0.22}$  \\
\enddata
\end{deluxetable*}

\subsection{Mass and Pressure Measurements}

With the best-fitting parameters for the gas density and temperature
profiles in hand, it is straightforward to determine the cluster
masses. The gas mass can be obtained by integrating the
density profile over a given volume,
\begin{equation}
  M_{\rm{gas}}(r)=4\pi \mu_{e} m_{p}\int n_{e}(r) r^2 dr,
  \label{Equ.: Gas mass}
\end{equation}
where $\mu_{e}m_{p}$ is the mean mass per electron.

The total mass can be obtained using Equation (\ref{Equ.: CC total
  mass}), with the $a_{\rm{cool}}$ parameter is set to 1 for
polytropic clusters. Moreover, the pressure profile can also be
obtained using Equation (\ref{Equ.: Pressure}). In Table \ref{Table:
  Cluster properties}, I present the gas and total cluster masses
enclosed within radii of $r_{2500}$ and $r_{500}$, corresponding to
densities 2500 and 500 times the critical density of the Universe at
the redshift of the cluster, respectively. Also listed in Table
\ref{Table: Cluster properties} are the pressure values obtained at
these radii. Measurements reported in this table take account of the
systematic uncertainties discussed in Section \ref{subsec:
  systematic}.

\begin{deluxetable*}{lCCCCCCCCC}
\tablenum{3}
\tablecaption{Cluster physical properties.\label{Table: Cluster properties}}
\tablewidth{0pt}
\tablehead{
\colhead{} &
\multicolumn4c{$\Delta = 2500$} &
\colhead{} &
\multicolumn4c{$\Delta = 500$} \\
\cline{2-5}
\cline{7-10}
\colhead{Cluster} &
\colhead{$r_{\Delta}$} &
\colhead{$M_{\rm gas}$} &
\colhead{$M_{\rm tot}$} &
\colhead{$P_e(r_\Delta)$} &
\colhead{} &
\colhead{$r_{\Delta}$} &
\colhead{$M_{\rm gas}$} &
\colhead{$M_{\rm tot}$} &
\colhead{$P_e(r_\Delta)$} \\
\colhead{} &
\colhead{(arcsec)} &
\colhead{$(10^{13}\,  M_{\odot})$} &
\colhead{$(10^{14}\,  M_{\odot})$} &
\colhead{($10^{-3}$ keV cm$^{-3}$)} &
\colhead{} &
\colhead{(arcsec)} &
\colhead{$(10^{13}\,  M_{\odot})$} &
\colhead{$(10^{14}\,  M_{\odot})$} &
\colhead{($10^{-3}$ keV cm$^{-3}$)}
}
\startdata
Abell 2218    & $195.5_{-3.7}^{+3.7}$ & $2.63_{-0.24}^{+0.42}$ &
$3.90_{-0.17}^{+0.27}$ & $2.31_{-0.32}^{+0.34}$ & & $362.4_{-6.3}^{+6.5}$ &
$5.96_{-0.52}^{+0.98}$ & $4.97_{-0.21}^{+0.33}$ & $0.53_{-0.13}^{+0.14}$ \\
Abell 1835    & $148.2_{-6.4}^{+7.7}$ & $4.17_{-0.36}^{+0.44}$ &
$4.63_{-0.45}^{+0.55}$ & $4.17_{-0.23}^{+0.25}$ & & $268.4_{-9.5}^{+12.4}$ &
$6.84_{-0.70}^{+0.91}$ & $5.54_{-0.61}^{+0.78}$ & $0.42_{-0.02}^{+0.03}$ \\
Abell 2050    & $253.5_{-20.8}^{+21.5}$ & $1.53_{-0.35}^{+0.31}$ &
$2.62_{-0.55}^{+0.48}$ & $2.41_{-0.44}^{+0.39}$ & &
$440.4_{-51.8}^{+39.7}$& $4.54_{-0.98}^{+0.63}$ &
$2.75_{-0.67}^{+0.44}$ & $0.74_{-0.02}^{+0.01}$ \\
Abell 1689    & $194.2_{-12.7}^{+13.6}$ & $2.61_{-0.47}^{+0.63}$ &
$3.61_{-0.69}^{+0.78}$ & $3.70_{-0.65}^{+0.85}$ & &
$351.4_{-23.1}^{+21.1}$ & $6.28_{-1.31}^{+1.61}$ &
$4.28_{-0.77}^{+0.79}$ & $0.93_{-0.17}^{+0.21}$ \\
MACS J0647.7+7015    & $63.9_{-3.1}^{+5.3}$ & $1.17_{-0.24}^{+0.25}$ &
$2.01_{-0.34}^{+0.36}$ & $3.59_{-0.60}^{+0.61}$ & &
$110.0_{-4.9}^{+9.3}$  & $2.75_{-0.44}^{+0.61}$ &
$2.68_{-0.48}^{+0.58}$ & $0.88_{-0.16}^{+0.18}$ \\
MACS J1423.8+2404    & $73.1_{-6.7}^{+7.1}$ & $2.38_{-0.38}^{+0.46}$ &
$2.59_{-0.53}^{+0.61}$ & $5.22_{-0.91}^{+0.94}$ & &
$132.7_{-11.6}^{+12.4}$ & $5.05_{-0.91}^{+0.93}$ &
$3.10_{-0.52}^{+0.56}$ & $0.97_{-0.18}^{+0.18}$ \\
RXC J2014.8-2430     & $165.2_{-5.9}^{+10.7}$ & $2.04_{-0.33}^{+0.48}$
& $1.59_{-0.18}^{+0.33}$ & $2.24_{-0.33}^{+0.41}$ & &
$286.1_{-11.2}^{+16.1}$ & $3.78_{-0.69}^{+0.87}$ &
$2.35_{-0.45}^{+0.58}$ & $0.48_{-0.09}^{+0.11}$
\enddata
\end{deluxetable*}

\subsection{Comparison With Previous Measurements}
\label{sec: mass comparison}

The mass measurements for Abell 1835, from this study, are
compared here with the results presented in \citet{landry2013}. I
estimate the cluster masses within the same angular radii as reported
in \citet{landry2013}. To allow a fair comparison of the cluster
masses, their uncertainties on $r_{2500}$ and $r_{500}$ are adopted.

Using the \citet{vikhlinin2006} model, \citet{landry2013} estimated
the gas mass of $4.60_{-0.25}^{+0.23} \times 10^{13}\, M_{\odot}$
and total mass of $4.74_{-0.61}^{+0.60} \times 10^{14}\, M_{\odot}$
within $r_{2500}=161.7_{-7.2}^{+6.5}$ arcsec. These values are consistent, at
the $1\sigma$ level, with the gas mass of $4.53_{-0.49}^{+0.55} \times
10^{13}\, M_{\odot}$ and total mass of $4.88_{-0.62}^{+0.75} \times
10^{14}\, M_{\odot}$ obtained in the current work for Abell
1835. Within $r_{500}=323.2_{-8.6}^{+8.6}$ arcsec, the predicted gas
and total masses of Abell 1835 by \citet{landry2013}, using the
\citet{vikhlinin2006} model, are $10.75_{-0.29}^{+0.28} \times
10^{13}\,M_{\odot}$ and $7.56_{-0.59}^{+0.62} \times 10^{14}\,
M_{\odot}$, respectively. These masses are about $26 \pm 6\%$ and $24
\pm 6\%$ larger than the corresponding masses of $7.95_{-0.79}^{+0.98}
\times 10^{13}\, M_{\odot}$ and $5.73_{-0.79}^{+0.99} \times 10^{14}\,
M_{\odot}$ given by this study within the same region.

Using the \citet{bulbul2010} model, the estimates obtained in
the \citet{landry2013} work for the gas and total masses are
$4.00_{-0.10}^{+0.10} \times 10^{13}\, M_{\odot}$ and
$3.33_{-0.20}^{+0.20} \times 10^{14}\, M_{\odot}$, respectively,
within $r_{2500}=143.8_{-2.9}^{+2.8}$ arcsec. The corresponding masses
predicted in this work are $4.10_{-0.31}^{+0.34} \times 10^{13}\,
M_{\odot}$ and $4.18_{-0.41}^{+0.50} \times 10^{14}\, M_{\odot}$. The
former value is consistent well with that reported by
\citet{landry2013}, whereas the total mass is about $20 \pm 5\%$ larger
than that given by these authors. Within $r_{500}=309.6_{-8.0}^{+8.2}$
arcsec, the estimated gas and total masses by \citet{landry2013} are
$10.36_{-0.27}^{+0.27} \times 10^{13}\,M_{\odot}$ and
$6.65_{-0.50}^{+0.54} \times 10^{14}\, M_{\odot}$, respectively. The
gas mass value is larger by about $24 \pm 4\%$ than the gas mass of
$7.85_{-0.67}^{+0.87} \times 10^{13}\,M_{\odot}$ derived by this
study, but the total mass is statistically consistent with the total
mass of $5.64_{-0.66}^{+0.74} \times 10^{14}\, M_{\odot}$ estimated by
the current study.

Overall, the mass measurements predicted from this study within the
$r_{2500}$ radii are in agreement with previous measurements. In the
cluster's outer regions, however, some of the mass measurements
are statistically inconsistent with previous measurements.  Such
discrepancies could be attributed to the choice of model for fitting
the X-ray data. \citet{landry2013} found that the choice of model may
introduce uncertainties of $6 \pm 6\%$ and $10 \pm 8\%$,
respectively, to the measurements of the cluster gas and total masses
at $r_{500}$. The temperature profile could be another possible source
for the discrepancy between these measurements. The temperature
profile used in this analysis is slightly different from that used in
\citet{landry2013}, and I estimated that the mass measurements for
Abell 1835 could be affected by adopting the current temperature
profile up to $\pm 5\%$ within the $r_{500}$ radius.

\begin{figure}
\plotone{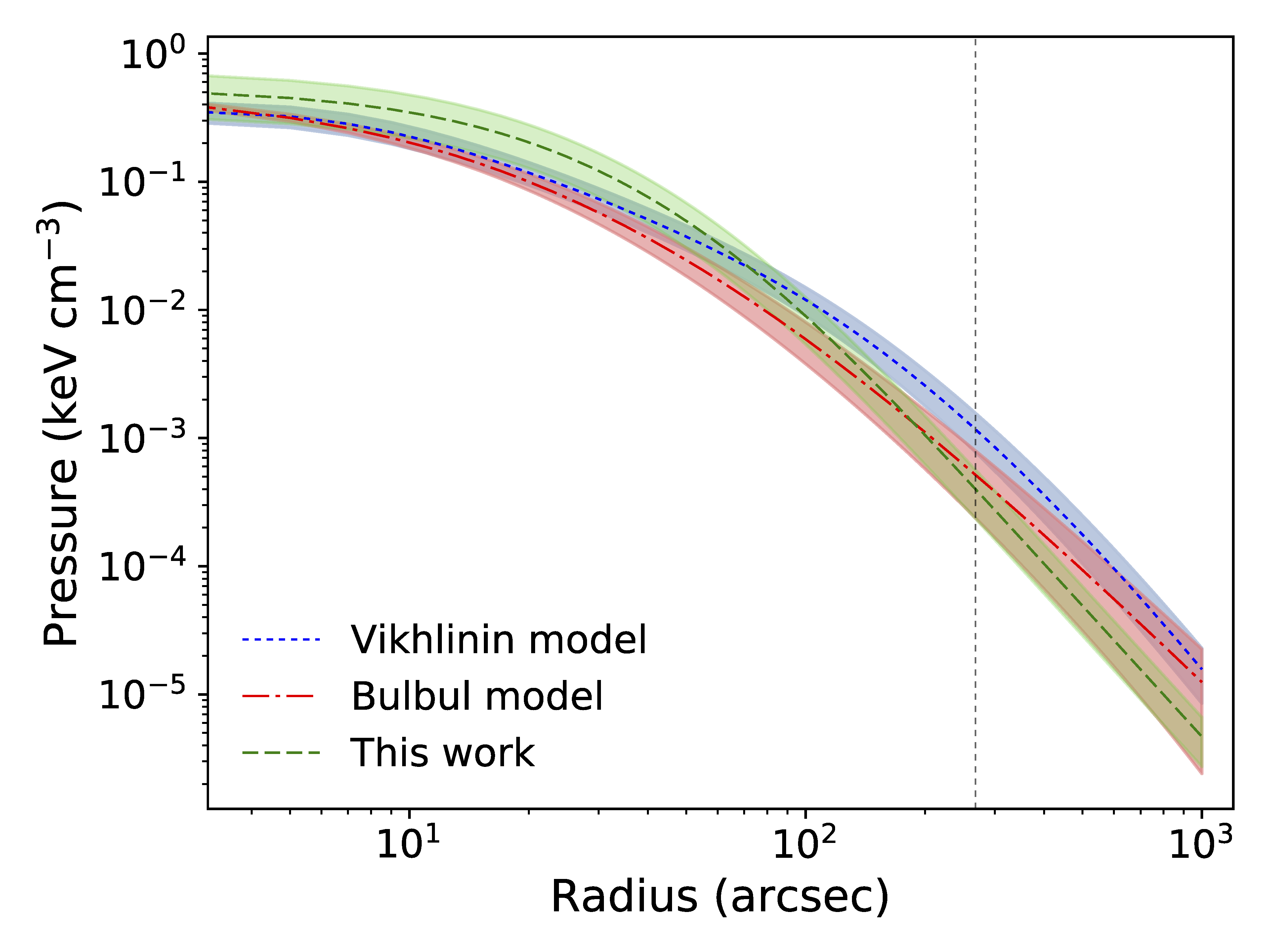}
\caption{Parameterized pressure profiles for Abell 1835 predicted from
  this work (Equation (\ref{Equ.: Pressure})) and \citet{landry2013}
  work. The lines are the best-fitting models, and the shadow regions
  are the 68.3$\%$ confidence intervals. The vertical-dashed line
  indicates the $r_{500}$ radius. The pressure profile seems robust in
  respect to different models.}
\label{Fig.: Pressure profiles}
\end{figure}

In addition to the mass comparison, the radial pressure distribution
for Abell 1835 is estimated using Equation (\ref{Equ.: Pressure}),
with the best-fitting parameters (Table \ref{Table: Best-fitting
  parameters}), and then compared with those predicted by
\citet{landry2013} using the \citet{vikhlinin2006} and
\citet{bulbul2010} models, and assuming $P_e(r)\, =n_e k_{\rm{B}}
T_e$. Figure \ref{Fig.: Pressure profiles} shows
the radial pressure profiles for Abell 1835 predicted from this work
and \citet{landry2013} work, associated with their 68.3$\%$ confidence
interval. This figure suggests that, despite the differences in the
temperature profiles adopted by the current work and
\citet{landry2013} work, the parameterized pressure profiles for Abell
1835 are more robust.

\subsection{Comparison With Previous Models}

The model proposed in the current work is compared with the
\citet{vikhlinin2006} model (Equations (\ref{Equ.: Density
  vikhlinin06}) and (\ref{Equ.: Temp vikhlinin06})) and the
\citet{bulbul2010} model (Equations (\ref{Equ.: Density bulbul}) and
(\ref{Equ.: Temp bulbul})) in order to test which model provides a
better fit to the X-ray data. For this purpose, the $\chi ^2$
statistic is used as a metric to compare these models. Table
\ref{Table: Comparison} shows the combined $\chi_{\rm{tot}}^2$ per
degree of freedom (reduced $\chi_{\rm{tot}}^2$) of the surface
brightness and temperature associated with each fit for all studied
galaxy clusters. As inferred by the reduced $\chi_{\rm{tot}}^2$, all
the models describe the data equally well. Similar to the \citet{bulbul2010}
model, however, the current model does not reproduce well the
temperature profile at intermediate and outer radii for some clusters,
such as Abell 1835 (Figure \ref{Fig.: cc clusters}).

\begin{deluxetable*}{lCCC}
\tablenum{4}
\tablecaption{Reduced $\chi_{\rm{tot}}^2$ values.\label{Table: Comparison}}
\tablewidth{0pt}
\tablehead{
\colhead{} &
\multicolumn3c{Reduced $\chi_{\rm{tot}}^2$} \\
\cline{2-4}
\colhead{Cluster} &
\colhead{This work} &
\colhead{\text{\citet{vikhlinin2006}}} &
\colhead{\text{\citet{bulbul2010}}}\\}
\startdata
Abell 2218        & 0.92  & 0.45  & 1.32  \\
Abell 1835        & 1.05  & 0.96 & 1.27 \\
Abell 2050        & 0.52 & 0.22  & 0.32 \\
Abell 1689        & 0.72  & 0.76  & 0.76 \\
MACS J0647.7+7015 & 0.54  & 1.12  & 0.69  \\
MACS J1423.8+2404 & 0.52  & 0.56 & 0.59  \\
RXC J2014.8-2430  & 1.12 & 1.16  & 1.13 \\
\enddata
\end{deluxetable*}

\section{DISCUSSION AND CONCLUSION}
\label{sec:discussion}

In this work, I present an analytical model for the density and
temperature profiles of galaxy clusters based on the assumption of
hydrostatic equilibrium in the cluster's gravitational potential. The
model represents a variation of the model proposed by
\citet{bulbul2010}. Here, the Einasto profile is adopted to model the
spatial-density distribution of dark matter halos instead of the
generalized NFW model used by \citet{bulbul2010}. This three-parameter
profile is initially combined with a polytropic equation of state. Then, a
cool-core correction is applied to the temperature profile and the gas
density profile is derived under the assumption that the pressure
profile is the same as in the polytropic case.

The model uses five parameters to describe the global
properties of the ICM gas, with three additional parameters
to describe the cluster's core region. The main advantage of this
model is the limited number of free parameters, which makes it simple
and robust. The robustness feature is particularly important when one
attempts to fit data that consist of few measurements. This can be
helpful, for example, in galaxy clusters, where the X-ray count
rate is low, particularly in the outskirts. Therefore, the proposed
model represents a practical improvement compared to the model
introduced by \citet{vikhlinin2006}, which has 17 free
parameters. From a computational point of view, it is also more
convenient to fit the observations to a model characterised by a few
parameters. Another feature of this new model is the weak degeneracies
between the best-fitting parameters, compared to the
\citet{vikhlinin2006} model (see Figures \ref{Fig.: Polytropic
  clusters} and \ref{Fig.: cc clusters}). This implies that more
precise estimates are obtained for parameters.

The model is tested observationally with the X-ray data for polytropic
and cool-core clusters. For most clusters, it is observed that the model
is able to accurately fit the radial distributions of the ICM
properties over the cluster's full radial range. It is also shown that
the model is essentially as good as that of \citet{vikhlinin2006} and
\citet{bulbul2010}, as indicated by the reduced $\chi^2$. Similar to
the \citet{bulbul2010} model, however, the model does not fit the
temperature profile well enough at intermediate and large radii for
some clusters.

Besides its application to model X-ray data, the model can be applied
to Sunyaev-Zel'dovich effect observations, making it useful for
various cosmological studies. The model can be used, for
example, to measure cosmic distances \citep{bonamente2006}, the
cluster pressure profiles \citep{bonamente2012}, and the gas mass
fractions \citep{planck2013}. Furthermore, the model can be used
to set up the initial conditions in cosmological numerical simulations,
since it provides a simple and accurate description of the properties
of the ICM plasma.

\section{ACKNOWLEDGEMENTS}

I am grateful to M. Birkinshaw for his useful comments, and I would
like to thank the anonymous referee for valuable comments and
suggestions which helped to improve the manuscript.

\end{document}